\documentclass{aa}  

\usepackage{graphicx}
\usepackage{txfonts}
\usepackage{lipsum}
\usepackage{subcaption}         % necessary for continued figures, example in section 3
\usepackage{lscape}             % to rotate a single page table, example in appendix.
\usepackage{placeins}           % useful with \FloatBarrier, to keep 
\begin{document}

%%%%%%%%%%%%%%%%%%%%%%%%%%%%%%%%%%%%%%%%
% if you use custom commands in your title,
% ensure to check your title when submitting!
%%%%%%%%%%%%%%%%%%%%%%%%%%%%%%%%%%%%%%%%
   \title{Revisiting orbital recurrence in the dimmings of Boyajian's star (KIC~8462852)}

   %\subtitle{A recurrent-fragment interpretation of Tabby\'s star}

%%%%%%%%%%%%%%%%%%%%%%%%%%%%%%%%%%%%%%%%
% Please separate each author with the \and command
%
% Use the \corrauth to provide the corresponding
% author address. It will be automatically inserted as 
% footnote in the PDF output.
%
% Please DO NOT include ORCIDs next to author names.
% Instead, please provide an active address for each coauthor:
% it will be automatically extracted by EDPS editorial system, 
% and co-authors will be be able to authenticate their ORCID.
%
% Only authenticated ORCIDs will be taken into account.
% ORCIDs included here will be removed.
%%%%%%%%%%%%%%%%%%%%%%%%%%%%%%%%%%%%%%%%

   \author{H. Socas-Navarro\inst{1}\corrauth{hector.socas@est-project.eu}        % use \corrauth for the corresponding author
        }

   \institute{European Solar Telescope Foundation, V\' \i a L\' actea S/N, La Laguna 38205, Tenerife, Spain}

   \date{Received September 10, 2026}

% \abstract{}{}{}{}{}
% 5 {} token are mandatory

 \abstract
% Context
{KIC~8462852 (Boyajian's star) displays deep, asymmetric and highly irregular
dimming events that show no simple repetition pattern. A symmetric
1.1\% transit detected by \emph{TESS} in 2019 has recently been
interpreted as a possible giant planet or brown dwarf companion, while the
irregular dips are generally attributed to transiting dust associated
with exocomets or planetesimal fragments.}
% Aims
{We investigate whether the absence of periodicity among individual
light-curve dips can nevertheless be compatible with orbital recurrence of the
broader dip families. In particular, we test a scenario in which a
major disruption event of a planetary body produces fragments with slightly
different orbital periods, causing successive returns of the fragments to
spread progressively in orbital phase.}
% Methods
{We assess two hypotheses in a Bayesian model comparison.
Under the null hypothesis, the centers of the detected dip families
are chronologically independent. Under the hypothesis of orbital recurrence,
we create a minimal model with two free parameters, a global period $P$ and 
a period-dispersion scale $\sigma_P$.
We marginalize over both parameters to compute the Bayes factor and
test the strength of the evidence
for the orbital recurrence model relative to the independent-chronology
hypothesis.}
% Results
{We obtain a Bayes factor $B_{10}=44.9$ in favor of orbital
recurrence, corresponding to ``very strong'' evidence on the
conventional Jeffreys scale. The maximum-likelihood solution is
$P_{\rm ML}=781.6$~d and $\sigma_{P,\rm ML}=80.0$~d.
The preferred period is close to the 776.1-d recurrence timescale
independently obtained if D800 and the 2019 \emph{TESS} event are
separated by four orbital revolutions. As an independent geometrical
consistency check not imposed by the timing model, interpreting
$P_{\rm ML}$ as the orbital period of the transiting body predicts
a transit duration of approximately 20.5~h, in close agreement with
the $\sim21$~h observed by \emph{TESS}.
}
% Conclusions
{Our analysis favors the orbital recurrence hypothesis, but it cannot
  be taken as definite proof because the Bayes factor obtained is
  conditional to the specific models explored. It is nevertheless a
  good argument to support this view, which we interpret as the result
  of a major disruption event occurring just before the D800 dip. The
  subsequent dimming episodes would then be recurrent transits of the
  fragments and associated dust.}


   \maketitle
   \nolinenumbers

%%%%%%%%%%%%%%%%%%%%%%%%%%%%%%%%%%%%%%%%%%%%%%%%%%%%%%%%%%%%%%

\section{Introduction}
\label{sec:introduction}

KIC~8462852 (Boyajian's star) is one of the most unusual objects observed by
the \emph{Kepler} mission. Its light curve contains a series of deep,
irregular dimming events whose depths, durations, and morphologies vary
strongly from one event to another \citep{boyajian2016}. The two most
prominent structures in the \emph{Kepler} data are the isolated event near
mission day 800 (hereafter D800), which reached a depth of approximately
15\%, and the much more complex sequence near day 1500 (D1500), which
extended over roughly 80~d and included a dip approaching 22\% in depth.
Subsequent ground-based monitoring revealed further, considerably shallower
dimming events in 2017, including the events commonly referred to as Elsie,
Celeste, Skara Brae, and Angkor \citep{boyajian2018}. These later events
typically had depths of a few percent and durations ranging from several
days to weeks. Similarly shallow and irregular dips were observed in 2018 (Caral-Supe and Evangeline, \citep{hitchcock2019}).

The chromatic nature of several of these events strongly supports extinction
by circumstellar dust \citep{deeg2018,boyajian2018}. Consequently, families
of exocomets or planetesimal fragments have long been among the leading
interpretations of the dips \citep{boyajian2016,bodman2016}. In such
scenarios, a population of bodies on similar orbits can generate complex
series of transits without requiring individual dips to repeat with the same
shape or depth. Several authors have also explored explicitly recurrent
interpretations of subsets of the observed events, including periodicities
of approximately 1570--1600~d and models involving orbiting rings, companions,
or associated debris populations \citep{sacco2018,bourne2018,kiefer2017,
ballesteros2018}.

Nevertheless, the lack of a simple repetition pattern has progressively
disfavored a periodic description of the phenomenon. A direct extrapolation from D800 and D1500 does not predict a
sequence of subsequent dips with comparable morphology, which go from a single event at D800 with nothing going on until nearly 800 days later, to the seemingly continuous chain of shallow dips in 2017 and 2018. Furthermore, the nature of the observed
events evolves from a single deep, very smooth and asymmetric structure at D800,
to the complex D1500 sequence, and later to much shallower and temporally
distributed events. All of these factors seem to argue against a simple periodic interpretation.

The recent analysis of \citet{madurga2026} adds an important new element to
this picture. Using \emph{TESS} observations, they identified a previously
unreported transit on 2019 September~3 with a duration of approximately 21~h
and a depth of about 1.1\%. Unlike the irregular dimming events observed by
\emph{Kepler}, the \emph{TESS} event is highly symmetric. Modeling of its
shape, together with radial-velocity observations and the available
photometric constraints, led them to favor an interpretation
in terms of a compact companion, possibly a giant planet or brown dwarf.
Within that interpretation, the allowed orbital periods are longer than
approximately 1030~d. This work treats the irregular dips as a separate, apparently
non-periodic phenomenon, most plausibly associated with dust-producing
exocomets or planetesimal fragments whose orbits are perturbed by the
massive companion.

Here we revisit a different possibility, namely that the absence of periodicity
among the \emph{individual} dips does not necessarily imply the absence of
orbital recurrence among the \emph{families} of dips. In this scenario, the occulting
material originated in a major, possibly catastrophic, disruption event of a parent body. The 
resulting fragments will end up with slightly different orbital periods and therefore would progressively separate in orbital
phase. Successive passages would then be expected to become increasingly
extended in time rather than to reproduce the original transit morphology.
At the same time, the dust content and optical depth of the fragment
population may evolve as material disperses and small grains are removed by
radiation pressure. Such an evolutionary picture can therefore produce
recurrent activity while accounting for the strong differences observed between
individual dimming events.

\begin{figure*}
    \centering
    \includegraphics[width=0.78\textwidth]{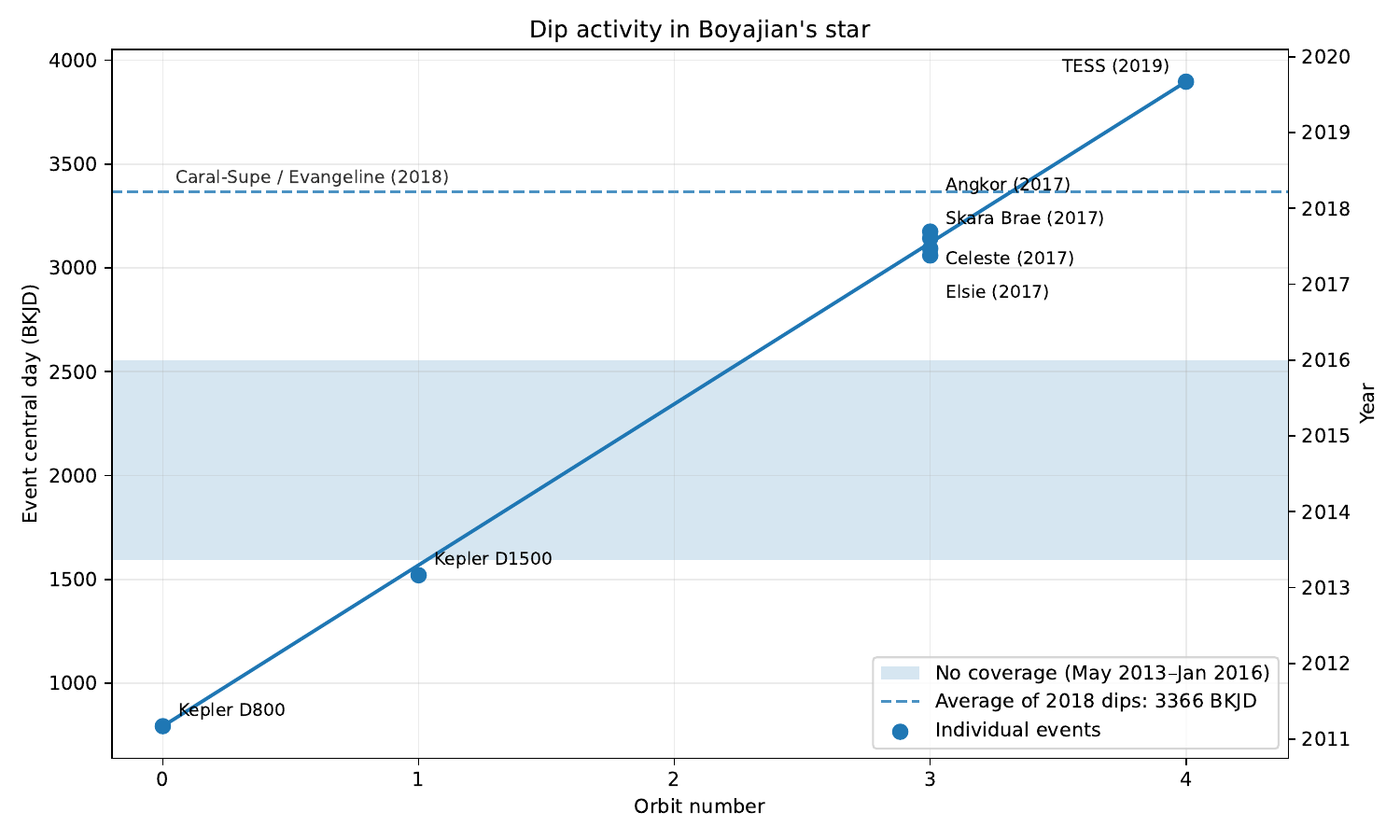}
    \caption{
    Temporal distribution of the main dimming families considered in this
    work. Individual events are shown together with representative epochs
    for D800, D1500, the 2017 activity, and the 2019 \emph{TESS} transit.
    The solid line illustrates the near-linear alignment of these
    representative epochs with recurrence number. The shaded region marks
    the interval without observational coverage, and the dashed
    horizontal line indicates the characteristic epoch of the 2018 activity.
    This graphical alignment is shown only as motivation; its statistical
    significance is assessed using the observing window and the models
    described in Sects.~\ref{sec:physical_scenario}--\ref{sec:bayes}.
    }
    \label{fig:family_timeline}
\end{figure*}

This possibility is particularly suggestive when the principal families are
considered only through their occurrence times. As illustrated in
Fig.~\ref{fig:family_timeline}, D800, D1500, the 2017 activity, and the 2019
\emph{TESS} transit define an approximately linear sequence when associated
with successive recurrent passages, while the 2018 activity occurs within
the temporally broadened region occupied by the later fragment population. 
According to this scenario, an intermediate return ($n=2$) should have
taken place between D1500 and the 2017 dip families.
This return would have fallen within the observational gap between the
end of \emph{Kepler}'s observations in 2013 and the beginning of
continuous ground-based monitoring in 2016.

The apparent alignment seen in the figure does not by itself constitute
evidence for periodicity. With a sparse and highly non-uniform observing
window, chance alignments can occur. Furthermore, the correspondence between the
observed families and the nominal recurrence epochs is not exact. Nevertheless, the
alignment motivates a quantitative comparison of the evidence for
orbital recurrence versus independent occurrence.

In this work, we use Bayesian model comparison to test whether the
chronology of the observed families is better described by a recurrent
fragment population (denoted $H_1$ and motivated by the \emph{Alderaan scenario} introduced below),
or by independent events that may
occur at any time (the null hypothesis, $H_0$). We begin by describing a minimal model
for both scenarios, and then
formulate both hypotheses taking into account the actual observing
window. We then proceed to compare their respective Bayesian evidences to derive the Bayes factor.

For this specified comparison, we obtain a Bayes factor
$B_{10}=44.9$ in favor of the orbital recurrent fragments hypothesis.
On the conventional Jeffreys scale, this value falls well within the
``very strong'' evidence range \citep{jeffreys1961}. We stress,
however, that such categories are descriptive guidelines rather
than detection thresholds. Following the recommendation of
\citet{kippingbenneke2025}, we do not translate the Bayes factor
into a frequentist $p$-value or an equivalent number of standard
deviations (sigmas), since no unique conversion between these quantities
exists and commonly used conversions can overestimate the
corresponding frequentist significance. We therefore report and
interpret the Bayes factor directly.

The maximum-likelihood recurrence period is close to 780~d, and
is therefore also close to the period obtained if D800 and the
2019 \emph{TESS} event are identified with passages separated by
four orbital revolutions. We emphasize that this result tests a
particular physical recurrence scenario rather than an unrestricted
search over all possible periodic patterns. Its definition, priors,
sensitivity to the observing window, and statistical interpretation
are developed in the following sections.

Remarkably, this result also passes an independent geometrical
consistency check. If the maximum-likelihood period,
$P_{\rm ML}=781.6$~d, is interpreted as the orbital period of the
compact body responsible for the 2019 \emph{TESS} event, then the
published stellar and transit parameters imply, for a circular orbit,
a transit duration of approximately 20.5~h. This agrees very closely
with the observed duration of $\sim21$~h
\citep{madurga2026}. The stellar mass, radius, impact parameter, and
transit geometry do not enter the recurrence-timing model, so this
agreement was not imposed by the Bayesian fit and provides an
independent physical consistency check of the Alderaan scenario. 
There is no a priori reason for a period inferred solely from the 
chronology of the dip families to reproduce the independently observed 
transit duration.

The paper is organized as follows. Section~\ref{sec:physical_scenario}
introduces the recurrent-fragment scenario and its observational
interpretation. Section~\ref{sec:models} defines the recurrent and independent
timing models and the observational selection function.
Section~\ref{sec:bayes} presents the Bayesian model comparison, and
Sect.~\ref{sec:discussion} discusses the physical interpretation,
limitations, and observational tests of the recurrence hypothesis.

\section{The physical scenario}
\label{sec:physical_scenario}

The hypothesis explored in this work is not that the individual dimming
events of Boyajian's star repeat periodically. Their very different depths,
durations, and morphologies clearly argue against such a simple picture.
Instead, we consider whether the observed \emph{families} of events may be
successive manifestations of a common population of fragments returning on
similar orbits.

We assume that a major disruption of a parent body occurred at, or before,
the epoch represented by D800. The disruption produced a population of
fragments with some spread in velocity, which results in slightly different orbital periods clustered around that of a surviving main
remnant. Even small differences in orbital period accumulate from one
revolution to the next. Consequently, an initially compact fragment
population progressively spreads in orbital phase, and the temporal interval
over which transits can occur increases with the number of completed
revolutions. We do not concern ourselves here with the nature of such an event, whether a cataclysmic collision, a tidal disruption or something else.

For convenience, we refer to this physical picture as the
\emph{Alderaan scenario}. It provides a natural way of reconciling
orbital recurrence with the strongly non-repeating appearance of the
observed dips. Its statistical implementation will be denoted by $H_1$
throughout the remainder of this work.
A recurrent passage need not
reproduce either the shape or the depth of the preceding one. Different
fragments may dominate the extinction at different epochs, dust production
may vary between fragments, and the dust itself evolves after release.
In particular, small grains are efficiently affected by radiation pressure,
so that a population of freshly generated dust tends to disperse over
time. We therefore expect recurrence, if present, to
be more readily identifiable in the timing of broad families of activity than
in the detailed morphology of individual dips.

The recurrent scenario considered here is motivated by the possibility
that D800 and the symmetric 2019 \emph{TESS} event correspond to passages of
the same surviving remnant. The morphology of the 2019 event, symmetric and shallow, motivated
\citet{madurga2026} to interpret it as the transit of a compact
companion, possibly a giant planet or brown dwarf. Here we explore the
possibility that this compact occulting body is the surviving
main remnant of the disruption responsible for the dip families.
If so, D800 and the TESS event can be used to define an approximate
recurrence timescale. Their separation is 3104.4~d. If four orbital
revolutions occurred between them (Fig.~\ref{fig:family_timeline}), the
corresponding period is 776.1~d. This choice predicts nominal return epochs
near the observed D1500 activity and the 2017 dimming events, as illustrated
in the Figure.

For this physical interpretation, we identify D800 with the reference passage
$n=0$, D1500 with $n=1$, the 2017 and 2018 activity with $n=3$, and the
2019 \emph{TESS} event with $n=4$. The corresponding nominal return epochs
for a period of 776.1~d are listed in Table~\ref{tab:topology}. As usual for \emph{Kepler} data, we use the Barycentric Kepler Julian Date (BKJD) as our time coordinate. It is defined as \(\mathrm{BKJD}=\mathrm{BJD}-2454833.0\), where BJD is the Barycentric Julian Date. Thus, \(\mathrm{BKJD}=0\) corresponds approximately to January 1, 2009 at noon UTC.

\begin{table}
\caption{Recurrence topology adopted for the long-period scenario.}
\label{tab:topology}
\centering
\begin{tabular}{ccc}
\hline\hline
Return $n$ & Nominal epoch (BKJD) & Associated activity \\
\hline
0 & 792.7  & D800 \\
1 & 1568.8 & D1500 \\
2 & 2344.9 & no assigned family \\
3 & 3121.1 & 2017 and 2018 \\
4 & 3897.2 & TESS 2019 \\
\hline
\end{tabular}
\end{table}

This topology is an explicit part of the Alderaan hypothesis tested below;
it is not a latent assignment optimized by the Bayesian calculation.
Nevertheless, within the long-period interpretation considered here, it is not
an arbitrary choice. If each observed family is required to remain associated
with a unique orbital return, that is, to lie within half an orbital period of
its nominal recurrence epoch, D1500 belongs to $n=1$, both the 2017 and 2018
activity belong to $n=3$, and the \emph{TESS} event belongs to $n=4$.
The $n=2$ passage falls in the interval without observational
coverage.

The physical scenario therefore makes two distinct predictions. First, the
centers of activity should remain associated with a common recurrence clock.
Second, the temporal region occupied by fragments may broaden from one return
to the next as differences in orbital period accumulate. In the following
section we translate these assumptions into a deliberately minimal timing
model. We do not attempt to model the number, depth, or detailed morphology
of individual dips which, in our scenario, are unpredictable (in practice) results 
from the disruption event.

\section{Timing models}
\label{sec:models}

\subsection{Representation of the observed families}
\label{sec:family_representation}

Dips occur clustered in families. Each observed family $k$ is represented by a time interval
\begin{equation}
F_k=[a_k,b_k],
\label{eq:family_interval}
\end{equation}
with center
\begin{equation}
c_k=\frac{a_k+b_k}{2},
\label{eq:family_center}
\end{equation}
and observed width
\begin{equation}
w_k=b_k-a_k.
\label{eq:family_width}
\end{equation}

The four families entering the statistical comparison are D1500,
the 2017 activity, the 2018 activity, and the 2019 \emph{TESS}
event. D800 is not included as a likelihood term because it defines
the phase reference, $t_0=792.7$~BKJD. The adopted family intervals
are listed in Table~\ref{tab:families}. Values in the table are rounded
for presentation.

\begin{table}
\caption{Observed family intervals used in the timing analysis.}
\label{tab:families}
\centering
\begin{tabular}{lrrr}
\hline\hline
Family & $a_k$ & $b_k$ & $w_k$ \\
       & (BKJD) & (BKJD) & (d) \\
\hline
D1500 & 1496.0 & 1568.5 & 72.5 \\
2017  & 3060.0 & 3173.0 & 113.0 \\
2018  & 3360.5 & 3370.0 & 9.5 \\
TESS  & 3896.7 & 3897.6 & 0.9 \\
\hline
\end{tabular}
\end{table}

The observed widths are conditioned upon in both hypotheses. In other
words, we do not attempt to infer the underlying distribution of family
widths. The widths nevertheless affect the calculation in two ways.
First, whether a family could have been detected depends on its temporal
extent and on the observing window. Second, under the recurrent model,
the complete observed family must fit within the temporal region allowed
for its orbital return.

The temporal coverage is highly non-uniform and must therefore be
included explicitly. We denote by
\begin{equation}
E_{\rm obs}(t)=
\begin{cases}
1, & \text{if suitable observations exist at time }t,\\
0, & \text{otherwise},
\end{cases}
\label{eq:eobs}
\end{equation}
the observing-window function. It includes the continuous
\emph{Kepler} coverage and the available post-\emph{Kepler} observing
epochs used in this work.

For convenience, let ${\cal E}$ denote the union of all time intervals
for which $E_{\rm obs}(t)=1$. For a family of width $w_k$, we define
$D_k$ as the set of all possible center times for which at least part
of such a family would overlap the observing window,

\begin{equation}
D_k =
\left\{
t:
\left[t-\frac{w_k}{2},t+\frac{w_k}{2}\right]
\cap {\cal E}
\neq \varnothing
\right\}.
\label{eq:Dk}
\end{equation}

Thus, $D_k$ does not represent the observations themselves. Rather, it
is the set of hypothetical \emph{family-center times} that could have
produced a detectable family of the observed width. For example, a wider family can remain detectable even if its center lies outside an observing interval, provided that part of the family overlaps that interval.

It is useful to introduce the indicator function

\begin{equation}
I_{0,k}(t)=
\begin{cases}
1, & t\in D_k,\\
0, & t\notin D_k.
\end{cases}
\label{eq:I0}
\end{equation}

The total amount of time over which the center of family $k$ could have
occurred and remained detectable is then

\begin{equation}
A_{0,k}
=
|D_k|
=
\int I_{0,k}(t)\,{\rm d}t.
\label{eq:A0}
\end{equation}
Here, $|D_k|$ denotes the measure, or total duration, of the set
$D_k$, rather than the absolute value of a number. Since $I_{0,k}$ is dimensionless, $A_{0,k}$ has units of time (in our case, days).

\subsection{Independent-family hypothesis ($H_0$)}
\label{sec:h0}

Under the null hypothesis, $H_0$, the occurrence times of the different
families are chronologically independent. Before conditioning on
detection, their centers are assumed to be uniformly distributed in
time over the analysis interval. The model therefore contains no
preferred recurrence timescale.

Once we condition on the fact that a family of width $w_k$ was detected,
its possible center is restricted to the detectable set $D_k$. Under
$H_0$, all times within this set are equally probable. The probability
density for the center of family $k$ is consequently
\begin{equation}
p_{0,k}(t\mid w_k,\mathrm{det},H_0)
=
\frac{I_{0,k}(t)}{A_{0,k}}.
\label{eq:h0_density}
\end{equation}

This expression is a probability \emph{density}, rather than a
probability. Since $A_{0,k}$ is measured in days,
$p_{0,k}(t)$ has units of ${\rm d}^{-1}$. For a sufficiently small
time interval ${\rm d}t$,
\begin{equation}
p_{0,k}(t\mid w_k,\mathrm{det},H_0)\,{\rm d}t
\end{equation}
is the probability that the center of a detected family of width $w_k$
lies between $t$ and $t+{\rm d}t$. By construction,
\begin{equation}
\int
p_{0,k}(t\mid w_k,\mathrm{det},H_0)\,
{\rm d}t
=1.
\label{eq:h0_normalization}
\end{equation}

Here and below, ``det'' denotes conditioning on the corresponding
family having been detected within the actual observing window.
Likewise, the observed family widths $\mathbf{w}$ are treated as
conditioned quantities rather than as observables to be generated by
either hypothesis. In other words, we do not expect our model to predict anything about the widths.
The observed center $c_k$ defined in Eq.~\eqref{eq:family_center} is one
particular realization of this random time. The likelihood contribution
of the observed family is therefore obtained by evaluating the density
at $t=c_k$,
\begin{equation}
p(c_k\mid w_k,\mathrm{det},H_0)
=
p_{0,k}(t=c_k\mid w_k,\mathrm{det},H_0)
=
\frac{1}{A_{0,k}},
\label{eq:h0_pdf}
\end{equation}
because the observed family necessarily satisfies $c_k\in D_k$.

The probability density in Eq.~\eqref{eq:h0_pdf} tells us how much
probability $H_0$ assigns to the particular center time that was
actually observed for family $k$. This is the basic meaning of a
\emph{likelihood}: it measures how compatible the observed data are with
a specified model. More precisely, for continuous data such as the
times considered here, the likelihood is the probability density
assigned by the model to the observed values. It should therefore not be
interpreted as the probability that the model itself is true.

For a single family, the likelihood contribution under $H_0$ is simply
\begin{equation}
{\cal L}_{0,k}
=
p(c_k\mid w_k,\mathrm{det},H_0)
=
\frac{1}{A_{0,k}}.
\label{eq:h0_single_likelihood}
\end{equation}

The full data set consists of the four observed family centers,
$\mathbf{c}=(c_1,c_2,c_3,c_4)$, with their corresponding observed
widths $\mathbf{w}=(w_1,w_2,w_3,w_4)$. Since the widths are conditioned
upon rather than modeled, the joint likelihood under $H_0$ is the
probability density of the observed centers conditional on these widths
and on their detection,
\begin{equation}
{\cal L}_0
=
p(\mathbf{c}\mid\mathbf{w},\mathrm{det},H_0)
=
p(c_1,c_2,c_3,c_4
\mid
w_1,w_2,w_3,w_4,\mathrm{det},H_0).
\label{eq:h0_joint_general}
\end{equation}

However, chronological independence is precisely one of the defining
assumptions of $H_0$. The joint density therefore factorizes into the
product of the individual densities,
\begin{equation}
p(\mathbf{c}\mid\mathbf{w},\mathrm{det},H_0)
=
\prod_k
p(c_k\mid w_k,\mathrm{det},H_0),
\label{eq:h0_independence}
\end{equation}
and hence
\begin{equation}
{\cal L}_0
=
\prod_k\frac{1}{A_{0,k}}.
\label{eq:h0_likelihood}
\end{equation}

Thus, in the null model $H_0$ the likelihood reduces to the familiar rule for
independent events: the joint probability density is obtained by
multiplying the probability densities of the individual observations.
After conditioning on the observed widths and on detection, $H_0$
contains no remaining continuous parameters.

Conditioning on the observed widths deliberately removes any information
carried by the width distribution itself from the model comparison.
Therefore, our statistical test asks only whether the observed family
centers are better explained by recurrent ($H_1$) or independent ($H_0$) timing, given
the widths that were actually observed. This choice potentially discards 
information favorable to the recurrent
interpretation, since the Alderaan scenario provides a physical mechanism
for the temporal spreading of successive families, whereas no corresponding
width evolution is produced under $H_0$. It is then a conservative choice in 
assessing the validity of $H_1$.

\subsection{Recurrent-fragment hypothesis ($H_1$)}
\label{sec:h1}

The recurrent hypothesis, $H_1$, is the statistical implementation of
the physical scenario described in Sect.~\ref{sec:physical_scenario}.
D800 defines the reference epoch $t_0$, and the nominal center of orbital
return $n$ is
\begin{equation}
\mu_n=t_0+nP,
\label{eq:mu_n}
\end{equation}
where $P$ is the recurrence period.

The fragments produced in the disruption are allowed to have slightly
different orbital periods. We describe their resulting temporal spread
using a single scale parameter, $\sigma_P$. In the minimal model adopted
here, the allowed temporal region for return $n$ is represented by a
top-hat window,
\begin{equation}
R_n(P,\sigma_P)
=
[\mu_n-n\sigma_P,\,
 \mu_n+n\sigma_P].
\label{eq:Rn}
\end{equation}

The temporal half-width therefore increases linearly with the number of
completed revolutions. The parameter $\sigma_P$ is a scale describing
how rapidly the fragments separate after each orbit (or equivalently, the spread in their periods).
Despite the notation, it is
not the standard deviation of an assumed Gaussian distribution; the
model has a temporal window in which the family must fit, given by
the interval $R_n$.

The recurrence topology specified in
Sect.~\ref{sec:physical_scenario} assigns each observed family $k$ to a
return number $n_k$. In the model considered here,
\begin{equation}
(n_{\rm D1500},n_{2017},n_{2018},n_{\rm TESS})
=
(1,3,3,4).
\end{equation}

A family of observed width $w_k$ and center $c_k$ occupies the interval
\begin{equation}
F_k=
\left[
c_k-\frac{w_k}{2},
c_k+\frac{w_k}{2}
\right].
\label{eq:family_from_center}
\end{equation}
For the observed family to be compatible with the recurrent model, this
entire interval must lie inside the recurrence window predicted for its
assigned return,
\begin{equation}
\left[
c_k-\frac{w_k}{2},
c_k+\frac{w_k}{2}
\right]
\subseteq
\left[
\mu_{n_k}-n_k\sigma_P,
\mu_{n_k}+n_k\sigma_P
\right].
\label{eq:family_inside_recurrence}
\end{equation}
This condition is equivalent to requiring both the left and right edges
of the family to remain within the recurrence window,
\begin{equation}
c_k-\frac{w_k}{2}
\geq
\mu_{n_k}-n_k\sigma_P,
\qquad
c_k+\frac{w_k}{2}
\leq
\mu_{n_k}+n_k\sigma_P.
\label{eq:family_edge_conditions}
\end{equation}
Solving these inequalities for the family center $c_k$ gives the range
of center times allowed by the recurrent model,
\begin{equation}
C_k(P,\sigma_P)
=
\left[
\mu_{n_k}-n_k\sigma_P+\frac{w_k}{2},
\,
\mu_{n_k}+n_k\sigma_P-\frac{w_k}{2}
\right].
\label{eq:Ck}
\end{equation}
Thus, $C_k$ is narrower than the full recurrence window
$R_{n_k}$ by $w_k/2$ at each edge. This simply reflects the requirement
that the complete family, rather than only its center, must fit inside
the predicted recurrence window.

It is not sufficient for an observed family to be contained within the
orbital recurrence window. It must also have occurred at a
time at which it could have been detected. Therefore, a valid center
must simultaneously satisfy
\begin{equation}
t\in C_k(P,\sigma_P)
\qquad\mathrm{and}\qquad
t\in D_k.
\end{equation}
The set of center times satisfying both conditions is consequently
\begin{equation}
S_k(P,\sigma_P)
=
C_k(P,\sigma_P)\cap D_k.
\label{eq:Sk}
\end{equation}

We define its indicator function as
\begin{equation}
I_{1,k}(t;P,\sigma_P)
=
\begin{cases}
1, & t\in S_k(P,\sigma_P),\\
0, & t\notin S_k(P,\sigma_P),
\end{cases}
\label{eq:I1}
\end{equation}
and its total duration as
\begin{equation}
A_{1,k}(P,\sigma_P)
=
|S_k(P,\sigma_P)|
=
\int I_{1,k}(t;P,\sigma_P)\,{\rm d}t.
\label{eq:A1}
\end{equation}
As for $A_{0,k}$, $A_{1,k}$ has units of days.

For fixed values of $P$ and $\sigma_P$, $H_1$ assigns equal probability
density to all center times belonging to $S_k$. The probability density
predicted for family $k$ is therefore
\begin{equation}
p_{1,k}
(t\mid P,\sigma_P,w_k,\mathrm{det},H_1)
=
\frac{I_{1,k}(t;P,\sigma_P)}
     {A_{1,k}(P,\sigma_P)}.
\label{eq:h1_density}
\end{equation}
As under $H_0$, this is a probability density in time, with units of
${\rm d}^{-1}$, and its integral over all possible center times is unity.

The likelihood contribution of the actual observed family is obtained by
evaluating this density at its observed center $t=c_k$,
\begin{equation}
{\cal L}_{1,k}(P,\sigma_P)
=
p_{1,k}
(c_k\mid P,\sigma_P,w_k,\mathrm{det},H_1).
\label{eq:h1_single_likelihood}
\end{equation}

Hence,
\begin{equation}
{\cal L}_{1,k}(P,\sigma_P)
=
\begin{cases}
\dfrac{1}{A_{1,k}(P,\sigma_P)},
&
c_k\in S_k(P,\sigma_P),
\\[2ex]
0,
&
c_k\notin S_k(P,\sigma_P).
\end{cases}
\label{eq:h1_pdf}
\end{equation}
For any fixed pair $(P,\sigma_P)$, the four observed family centers are
then treated as conditionally independent realizations of their assigned
recurrence windows. Their joint likelihood is therefore
\begin{equation}
{\cal L}_1(P,\sigma_P)
=
\prod_k
{\cal L}_{1,k}(P,\sigma_P).
\label{eq:h1_likelihood}
\end{equation}

Equations~\eqref{eq:h0_density} and \eqref{eq:h1_density} highlight the
basic difference between the two hypotheses. Under $H_0$, the center of
a detected family may occur anywhere in the full detectable set $D_k$.
Under $H_1$, it is restricted to the subset of detectable times that is
also compatible with the appropriate orbital return. The Bayesian
comparison in Sect.~\ref{sec:bayes} quantifies whether the concentration
of the observed centers within these recurrent regions compensates for
the additional freedom introduced by the parameters $P$ and $\sigma_P$.

For compactness, in the remainder of the paper we generally omit the
explicit time argument and write probabilities evaluated directly at the
observed centers $c_k$.

\subsection{Orbital-cell consistency and empty returns}
\label{sec:consistency}

The fixed topology specified in Sect.~\ref{sec:physical_scenario} assigns
each family to a particular return number $n_k$. However, this assignment
can only remain meaningful for values of $P$ for which the observed
family actually lies closer in orbital phase to its assigned return than
to either neighboring return.

For a trial period $P$, we define the orbital cell of return $n$ as the
interval bounded by the midpoints between the neighboring nominal
returns,
\begin{equation}
V_n(P)
=
\left[
t_0+\left(n-\frac{1}{2}\right)P,
\,
t_0+\left(n+\frac{1}{2}\right)P
\right].
\label{eq:orbital_cell}
\end{equation}

For family $k$ to remain consistently assigned to return $n_k$, the
complete observed family interval must satisfy
\begin{equation}
F_k
\subseteq
V_{n_k}(P).
\label{eq:cell_condition}
\end{equation}
If this condition fails for any family, that value of $P$ is inconsistent
with the topology defining $H_1$, and we therefore set
\begin{equation}
{\cal L}_1(P,\sigma_P)=0.
\end{equation}
The orbital-cell boundary is used only as a hard consistency check. It is
not used to truncate $C_k$, $S_k$, or $A_{1,k}$, because doing so would
artificially increase the likelihood whenever a cell boundary happened
to lie close to an observed family. This requirement therefore prevents
the fixed topology from being retained at periods for which a family
would instead belong to a neighboring orbital return, without affecting
the likelihood normalization within its assigned return.

We impose one additional constraint using the continuous \emph{Kepler}
coverage. We assume that an orbital return that occurs during the
\emph{Kepler} observations should have been detected. For any
post-D800 return, we consider a minimum selected family width of
\begin{equation}
w_{\rm min}=1~{\rm d}.
\end{equation}

If, for a given $(P,\sigma_P)$, the complete range in which such a family
could occur is contained within the continuous \emph{Kepler} coverage,
but no observed family is associated with that return, we assign zero likelihood to it:
\begin{equation}
{\cal L}_1(P,\sigma_P)=0.
\end{equation}
In other words, any combination of $(P,\sigma_P)$ that predicts dips during the \emph{Kepler} series that were not actually observed is considered incompatible with the data. The condition is deliberately strict: a return is vetoed only when the
observations would have guaranteed coverage of the entire allowed region. This is useful, e.g. to reject short orbital periods that would have produced families between D800 and D1500. Outside the \emph{Kepler} domain we are not so strict, because ground-based observations can suffer of bad weather, inhomogeneous conditions and other effects that might hinder the detection of dips that actually happened. Thus, no empty-return penalty is applied outside the \emph{Kepler} series.

The calculation otherwise remains conditional on the observed number of
families. In particular, we do not introduce a probabilistic model for
the number of families generated per return, nor for the number, depth,
or detailed timing of individual dips within a family.

\section{Bayesian hypothesis comparison}
\label{sec:bayes}

\subsection{Priors and evidences}

The Bayesian evidence for a hypothesis $H$ is
\begin{equation}
Z_H = p({\cal D}\mid H)
=
\int
{\cal L}(\boldsymbol{\theta})
\pi(\boldsymbol{\theta}\mid H)
\,{\rm d}\boldsymbol{\theta},
\label{eq:evidence}
\end{equation}
where ${\cal D}$ denotes the observed family timings,
$\boldsymbol{\theta}$ the model parameters, ${\cal L}$ the likelihood, and
$\pi$ the prior density.

For $H_0$, after conditioning on family widths and detectability, there are no
remaining free parameters. The evidence is therefore simply
\begin{equation}
Z_0 =
\prod_k \frac{1}{A_{0,k}}.
\end{equation}

For $H_1$, the two free parameters are $P$ and $\sigma_P$. We need to establish the priors for these two quantities. For $P$, we adopted
\begin{equation}
P \sim {\cal U}(120,1200)\ {\rm d} \, .
\end{equation}
The $\sim$ symbol here means that samples are drawn from a given
distribution, in this case uniform ${\cal U}$ between the specified
values. The short end of the period range is already in the region
ruled out by \emph{Kepler}, whereas the long end is limited by the
duration of the time series.

For $\sigma_P$ we chose a log-uniform distribution:
\begin{equation}
\sigma_P
\sim
{\rm LogUniform}(1,200)\ {\rm d}.
\end{equation}

Thus,
\begin{equation}
\pi(P)=\frac{1}{1080}
\end{equation}
within the adopted period interval, while
\begin{equation}
\pi(\sigma_P)
=
\frac{1}
{\sigma_P\ln(200)}
\end{equation}
for $1 \le \sigma_P \le 200$~d. The evidence for recurrence is
\begin{equation}
Z_1 =
\int_{120}^{1200}
\int_{1}^{200}
{\cal L}_1(P,\sigma_P)
\pi(P)\pi(\sigma_P)
\,{\rm d}\sigma_P\,{\rm d}P.
\label{eq:z1}
\end{equation}

Because the recurrent model has only two free parameters, we evaluate this
integral directly using two-dimensional Gauss--Legendre quadrature rather
than stochastic sampling. The integration is performed in the cumulative
prior coordinates, so that the quadrature directly averages the likelihood
over the normalized prior volume.

The Bayes factor comparing recurrence with independent occurrence is
\begin{equation}
B_{10}
=
\frac{Z_1}{Z_0},
\end{equation}
or equivalently
\begin{equation}
\ln B_{10}
=
\ln Z_1-\ln Z_0.
\end{equation}

\subsection{Results}

For the observing window and family intervals defined above, the effective
detectable-center measures under $H_0$ are
\begin{equation}
\begin{split}
A_{0,\mathrm{D1500}} &= 1911.300~{\rm d},\\
A_{0,2017} &= 1992.260~{\rm d},\\
A_{0,2018} &= 1722.760~{\rm d},\\
A_{0,\mathrm{TESS}} &= 1468.135~{\rm d}.
\end{split}
\end{equation}

These give
\begin{equation}
%\ln Z_0 = -29.895995.
\ln Z_0 = -29.89.
\end{equation}

Using $200\times200$ quadrature nodes for the recurrent model, we obtain
\begin{equation}
%\ln Z_1 = -26.092580,
\ln Z_1 = -26.09,
\end{equation}
and hence
\begin{equation}
\ln B_{10}=3.80
\end{equation}
or
\begin{equation}
B_{10}=44.9.
\end{equation}

Thus, under the explicitly stated conditioning and recurrence topology,
the observed family timings are approximately 45 times more probable under
the orbital recurrence model than under the null hypothesis.

The marginal posterior for the recurrence period is
\begin{equation}
P =
800.9^{+41.1}_{-28.4}\ {\rm d},
\end{equation}
where the quoted limits correspond to the 16th and 84th percentiles.
For the dispersion parameter we obtain
\begin{equation}
\sigma_P =
131.2^{+39.7}_{-32.2}\ {\rm d}.
\end{equation}

The maximum-likelihood point is distinct from the posterior median. The best
quadrature point is
\begin{equation}
P_{\rm ML}=781.63~{\rm d},
\end{equation}

\begin{equation}
\sigma_{P,\rm ML}=79.99~{\rm d},
\end{equation}
with
\begin{equation}
%\ln {\cal L}_{1,\rm max}=-20.5322.
\ln {\cal L}_{1,\rm max}=-20.53.
\end{equation}

At this point, the nominal return centers for $n=1$, 3, and 4 are
approximately 1574.4, 3137.6, and 3919.3~BKJD, respectively. D1500, the 2017
and 2018 families, and the \emph{TESS} event all satisfy the corresponding
orbital-cell and recurrence-window constraints.

The maximum-likelihood period is particularly noteworthy because it is close
to the value obtained independently by identifying D800 and the
\emph{TESS} event with passages separated by four revolutions,
\begin{equation}
P_{\rm D800-TESS}=776.106~{\rm d}.
\end{equation}

The difference between the maximum likelihood and the posterior median is
expected. The likelihood reaches its highest value near
$(P,\sigma_P)\simeq(782,80)$~d, whereas a broader region at somewhat larger
$\sigma_P$ occupies considerably more prior volume and therefore shifts the
marginal posterior.

As a diagnostic, the maximum log-likelihood improvement relative to $H_0$ is
\begin{equation}
\Delta\ln{\cal L}_{\rm max}
=
%9.364,
9.36,
\end{equation}
corresponding to a maximum-likelihood ratio of approximately
$1.2\times10^4$. This value is not the Bayes factor: after marginalizing over
the full prior ranges in $P$ and $\sigma_P$, the evidence ratio is reduced to
$B_{10}\simeq45$. 

We tested the numerical stability of the evidence using between 160 and 320
quadrature nodes per dimension. The resulting Bayes factors remain close to
$B_{10}\simeq45$, with variations at the few-percent level, indicating that
the numerical integration uncertainty is small compared with the modeling
assumptions.

The interpretation of this result is necessarily conditional. The Bayes
factor compares the specific long-period recurrent scenario defined in
Sect.~\ref{sec:physical_scenario} with chronologically independent family
times. It conditions on the observed family widths, the observed family
count, D800 as the phase reference, and the adopted recurrence topology. It
does not constitute an unrestricted search over every possible periodic
assignment. The consequences of these assumptions and the sensitivity of the
result to the physical model are discussed in Sect.~\ref{sec:discussion}.

\section{Discussion}
\label{sec:discussion}

\subsection{Evidence for orbital recurrence}

The main result of this work is that the chronology of the observed
dimming families is better described by the recurrent-fragment model
defined in Sect.~\ref{sec:physical_scenario} than by the hypothesis
that the family centers are chronologically independent. After
marginalizing over the recurrence period and the fragment-period
dispersion, we obtain a Bayes factor $B_{10}=44.9$.

On the conventional Jeffreys scale, Bayes factors between
approximately 32 and 100 are usually described as ``very strong''
evidence \citep{jeffreys1961}. According to this terminology, the
observed family timings provide very strong evidence in
favor of the recurrent model over the particular null hypothesis
considered here. We stress, however, that these qualitative labels
are descriptive guidelines rather than detection thresholds.

A Bayes factor is a measure of the relative support provided by the
data to two explicitly specified models. In the present case,
$B_{10}=44.9$ means that, after averaging over the stated priors, the
marginal probability density of the observed family timings is 44.9
times larger under $H_1$ than under $H_0$. It is not a frequentist
false-alarm probability, nor does it directly represent the posterior
probability that $H_1$ is true.

For the same reason, we do not convert $B_{10}$ into a $p$-value or
an equivalent number of standard deviations. As discussed by
\citet{kippingbenneke2025}, there is no unique conversion between
Bayes factors and frequentist significance, and commonly used
transformations can lead to an overstatement of the latter. We
therefore report and interpret the Bayes factor directly.

The maximum-likelihood solution provides an additional noteworthy
result. We find $P_{\rm ML}=781.6$~d and
$\sigma_{P,\rm ML}=80.0$~d. The preferred recurrence period is close
to the 776.1-d period obtained independently by assuming that D800
and the 2019 \emph{TESS} event correspond to passages of the same
surviving body separated by four revolutions. This value is not
imposed by the Bayesian calculation: $P$ is allowed to vary over the
full prior range from 120 to 1200~d. The proximity of these two
timescales therefore results from the simultaneous requirement to
accommodate D1500 and the 2017--2018 activity.

These results do not demonstrate that the dimming episodes of
Boyajian's star are caused by orbital recurrence. They show instead
that recurrence is a quantitatively viable and, within the specific
comparison performed here, favored explanation of their temporal
organization. In particular, the absence of periodicity among the
individual dips should not by itself be interpreted as evidence
against recurrence of the underlying fragment population.

\subsection{Physical interpretation}

The Alderaan scenario provides a simple physical interpretation of
the observed evolution. A major disruption produces a surviving
parent body or main remnant together with a population of smaller
fragments and dust. Immediately after disruption, the fragments
occupy neighboring orbital phases. Small differences in orbital
period then accumulate from one revolution to the next, progressively
spreading the population in time.

The observed light curves are qualitatively compatible with such an
evolution. D800 is a comparatively compact event with little internal
temporal structure. D1500 consists of a much more extended and complex
sequence of deep occultations. The 2017--2018 activity is distributed
over an even longer interval and consists predominantly of
considerably shallower dips. In this picture, the increasing temporal
extent is associated with orbital shear among the fragments, whereas
the decreasing prominence of the occultations may reflect the
evolution of their dust content. Grains may be efficiently removed by
radiation pressure (especially the smaller ones) or be reabsorbed
gravitationally, and the dust generated during the original disruption
or by subsequent fragment activity need not survive indefinitely.

We stress that the statistical model presented in this work does not
use either dip depth or morphology. These properties therefore do not
contribute to the Bayes factor. Their qualitative evolution instead
provides independent physical support for the recurrent-fragment
interpretation. Neither do we require dip depth to decrease
monotonically with time: individual fragments may produce fresh dust,
so substantial stochastic variability is expected within the overall
evolution.

The 2019 \emph{TESS} transit occupies a particularly interesting
position in this picture. Its shallow and symmetric morphology led
\citet{madurga2026} to favor a compact companion, possibly in the
giant-planet or brown-dwarf regime. In the interpretation explored
here, the compact occulting body could instead be the surviving main
remnant of the disruption, while the irregular dips arise from
material progressively separating from it.

This interpretation also passes an independent geometrical
consistency check. For a circular orbit, the total transit duration
is related to the scaled semimajor axis by
\begin{equation}
T_{14} =
\frac{P}{\pi}
\arcsin\left[
\frac{\sqrt{(1+k)^2-b^2}}
{\sqrt{(a/R_\star)^2-b^2}}
\right],
\end{equation}
where $k=R_c/R_\star$ and $b$ is the impact parameter.
Madurga-Favieres et al. (2026) obtain
$k=0.111\pm0.004$ and
$b=0.76^{+0.05}_{-0.04}$ for the 2019 transit, while adopting
$R_\star=1.60^{+0.07}_{-0.08}\,R_\odot$ and
$M_\star=1.19^{+0.20}_{-0.15}\,M_\odot$.

Using our maximum-likelihood period, $P=781.6$~d, together with
the nominal stellar mass and radius and the published transit
geometry, gives a circular-orbit semimajor axis of approximately
$1.76$~AU and a predicted transit duration of approximately
$20.5$~h. This is remarkably close to the observed duration of
$\sim21$~h.

Conversely, fixing $P=781.6$~d and $T_{14}=21$~h and using the
published values of $b$ and $k$ gives
$a/R_\star\simeq230$ and a stellar density of approximately
$3.8\times10^2$~kg~m$^{-3}$. For
$R_\star=1.60\,R_\odot$, this corresponds to
$M_\star\simeq1.10\,M_\odot$, well within the published uncertainty
on the stellar mass. Thus, the duration of the \emph{TESS} transit
provides an independent consistency check on the $\sim780$-d
recurrence timescale inferred from the dip chronology.

We emphasize that this calculation is meant as an independent test
of the physical scenario, and not yet an new
precision measurement of the stellar mass. The quoted 21-h duration
is approximate, and the inferred stellar density is correlated with
the impact parameter, radius ratio, limb darkening, and, if allowed,
orbital eccentricity. If the recurrence hypothesis is adopted, a
dedicated refit of the \emph{TESS} transit with the orbital period
fixed near 780~d could provide an independent constraint on
$\rho_\star$, which could then be combined with an external stellar
radius to refine the stellar mass.

\subsection{The timing of the disruption}

An intriguing aspect of the Alderaan scenario is that D800 may
correspond not merely to an arbitrary reference passage, but to an
epoch very close to the disruption itself. D800 is narrow compared
with the subsequent families and displays little internal temporal
structure. If this width is interpreted as an upper limit on the
initial phase spread of the newly produced fragment population, the
disruption must have occurred shortly before the D800 conjunction.

The scale of this effect can be estimated directly from the
maximum-likelihood solution. The inferred fragment-spreading scale is
approximately $\sigma_P\simeq80$~d per $P\simeq780$-d revolution.
Since $\sigma_P$ represents the half-width of the allowed temporal
distribution after one revolution, a D800 total width of order 2~d
corresponds to an initial half-width of order 1~d. If the phase
spreading is extrapolated approximately linearly back toward the
disruption epoch, the elapsed time $\Delta t$ is therefore of order

\begin{equation}
\Delta t
\sim
P\,\frac{1~{\rm d}}{\sigma_P}
\sim
780~{\rm d}\,\frac{1}{80}
\sim
10~{\rm d}.
\end{equation}

Thus, within the simplest version of the scenario, the catastrophic
disruption would have taken place only of order ten days before the
D800 transit.

This estimate should not be interpreted as a precise measurement of
the disruption epoch. The observed width of D800 includes the
geometrical duration of the occultation and the spatial extent of the
occulting material, and need not be produced entirely by orbital
dispersion. Moreover, our statistical model describes the accumulated
spread after complete orbital revolutions and does not explicitly
model the detailed dynamics during the first days following
disruption. The calculation should therefore be regarded as an
order-of-magnitude argument. Nevertheless, it gives a physical reason
why D800 is a particularly natural choice for the $n=0$ epoch: at
this time the parent body has a well-defined orbital phase and the
fragment population appears to have accumulated very little phase
dispersion.

If correct, this interpretation implies a remarkable observational
circumstance. The four-year \emph{Kepler} mission would have observed
an apparently mature planetary system within only days to weeks of a
major disruption event. Furthermore, the disruption would have
occurred close to the particular orbital phase at which the newly
produced material subsequently crossed our line of sight. For an
orbital period of approximately 780~d, an interval of order ten days
represents only a small fraction of an orbital cycle.

The transient nature of such a catastrophic event in an apparently
mature system has previously been recognized as a difficulty for
disruption-based interpretations of KIC~8462852
\citep[e.g.][]{boyajian2016,marengo2015,metzger2017}. We do not
attempt to assign a probability to having observed this coincidence.
The relevant disruption rate, transit geometry, and observational
selection effects are not sufficiently well constrained for such a
calculation. Nevertheless, the coincidence is a striking consequence
of the simplest recurrent-fragment interpretation and, at the same
time, provides the unusually well-defined initial condition that
makes the subsequent recurrence pattern testable.

\subsection{Limitations of the model}

The strength of the evidence must be understood in the context of the
specific hypotheses that were compared. H1 is not an unrestricted
periodicity search. It assumes the long-period topology
\begin{equation}
(n_{\rm D1500},n_{2017},n_{2018},n_{\rm TESS})
=
(1,3,3,4)
\end{equation}
described in Sect.~\ref{sec:physical_scenario}.  This topology is
physically motivated and is the unique assignment associated with the
approximately 800-d recurrence considered here when each observed
family is required to belong to a single orbital cell.  Nevertheless,
the resulting Bayes factor remains conditional on this choice.

Similarly, H0 represents chronologically independent family times.
It is intended as the simplest realization of the hypothesis that the
observed families have no common recurrence clock. It does not
represent every conceivable aperiodic physical mechanism. A more
complex null hypothesis in which unrelated activity itself occurs in
temporally clustered episodes would constitute a different model and
could lead to a different Bayes factor.

The analysis also conditions on the observed number and widths of the
families. We therefore do not use as statistical evidence either the
apparent increase in temporal extent of successive families or the
evolution of their dip depths. This was a deliberate conservative
choice that keeps the model comparison focused on timing. A complete
generative model would instead attempt to predict the number of
fragments, their period distribution, the number of detectable
families per return, their optical depths, and their dust evolution.
The available observations do not appear sufficient to constrain such
a model without introducing substantially more assumptions.

Our treatment of detectability is also necessarily approximate. The
continuous \emph{Kepler} observations allow a strong constraint on
missing returns, whereas the post-\emph{Kepler} observing window is
sparse and heterogeneous. Weather, cadence, photometric precision,
and dip depth all influence whether an event would actually have been
detected. We therefore apply the empty-return veto only when the
continuous \emph{Kepler} coverage would have guaranteed observation
of the complete allowed region.

Finally, the numerical value of a Bayesian evidence depends on its
priors. Our choices for $P$ and $\sigma_P$ form part of the
definition of H1 and are stated explicitly in
Sect.~\ref{sec:bayes}. The value $B_{10}=44.9$ should consequently
be interpreted as the evidence for the particular hypotheses, priors,
family definitions, and observational selection function considered
here, rather than as a prior-independent property of the data.

\subsection{Future evolution and observational tests}

The inferred fragment dispersion implies that distinct recurrent
families cannot remain recognizable indefinitely. At the
maximum-likelihood solution, $\sigma_P\simeq80$~d, so that by the
fifth return the characteristic half-width $n\sigma_P$ is already of
order 400~d, comparable to half of the $\sim780$-d orbital period.
The populations associated with neighboring returns should therefore
begin to overlap substantially in orbital phase.

The model thus predicts that after the 2019 epoch the concept of
discrete, well-separated families should progressively lose its
meaning. If the reservoir of small dust grains also decreases with
time, later occultations may become shallower and less structured.
The isolated and symmetric \emph{TESS} transit is qualitatively
consistent with such an evolution, although our model does not imply
that all dust must have disappeared or that further dipping activity
cannot occur.

Continued photometric and radial-velocity monitoring can provide
particularly useful tests. A secure orbital period for the object
responsible for the 2019 transit would directly test its proposed
association with D800. Likewise, the temporal distribution, depth,
and chromaticity of any future dipping episodes can test whether the
occulting material continues to evolve as an orbitally sheared
fragment population.

In summary, the present analysis does not establish orbital
recurrence as the origin of the remarkable dimming behavior of
Boyajian's star. It does show that the observed family timings contain
substantially more support for the specified recurrent-fragment
scenario than for chronologically independent occurrence. Combined
with the natural phase spreading expected from a disrupted
population, the morphological evolution of the observed dips, and the
possible association of D800 with an epoch very close to the original
disruption, this makes orbital recurrence an attractive and physically
grounded hypothesis that deserves to remain under consideration.

\begin{acknowledgements}
This research made use of NASA's Astrophysics Data System
Bibliographic Services. The analysis codes and figures made use of
NumPy \citep{harris2020}, Matplotlib \citep{hunter2007}, and IPython
\citep{perez2007}.  The author used a large language model (OpenAI's
GPT-5.6 Sol) for language editing, discussions and assistance with
code development. All scientific choices, source verification, code
execution, interpretation, and final text were conducted or reviewed
by the author, who takes full responsibility for the content of the
manuscript.
\end{acknowledgements}

\bibliographystyle{bibtex/aa}
\bibliography{bibtex/references}

\end{document}